\documentclass{ws-rv9x6}
\usepackage{subfigure}   
\usepackage{ws-rv-thm}   
\usepackage{ws-rv-har}   
\usepackage{txfonts}
\makeindex

\newcommand{\ltsima}{$\; \buildrel < \over \sim \;$}
\newcommand{\lsim}{\lower.5ex\hbox{\ltsima}}
\newcommand{\gtsima}{$\; \buildrel > \over \sim \;$}
\newcommand{\gsim}{\lower.5ex\hbox{\gtsima}}
\newcommand{\bra}{\langle}
\newcommand{\ket}{\rangle}

\newcommand{\ci}{\mathrm{i}}
\newcommand{\dd}{\mathrm{d}}
\newcommand{\veck}{\mathbf{k}}
\newcommand{\vecx}{\mathbf{x}}
\newcommand{\vecr}{\mathbf{r}}
\newcommand{\vecv}{\mathbf{\upsilon}}
\newcommand{\vecw}{\mathbf{\omega}}

\newcommand{\vecq}{\mathbf{q}}

\begin{document}

\setcounter{chapter}{1}
\chapter[cosmic structure formation]{Formation of the First Black holes:\\ Formation and evolution of the cosmic large-scale structure$^1$}\label{bjoern_chapter}

\author[B.M. Sch{\"a}fer]{Bj{\"o}rn Malte Sch{\"a}fer}

\address{Zentrum f{\"u}r Astronomie der Universit{\"a}t Heidelberg,\\ 
Astronomisches Recheninstitut\\
Philosophenweg 12, 69120 Heidelberg, Germany\\
bjoern.malte.schaefer@uni-heidelberg.de}

\footnotetext{$^1$ Preprint~of~a~review volume chapter to be published in Latif, M., \& Schleicher, D.R.G., ''Formation and evolution of the cosmic large-scale structure'', Formation of the First Black Holes, 2018, pages 22-44, \textcopyright Copyright World Scientific Publishing Company, https://www.worldscientific.com/worldscibooks/10.1142/10652 }.

\begin{abstract}
\end{abstract}
\body

\setcounter{page}{22}

\section{Introduction}\label{bjoern_intro}
The formation of cosmic structures is a fluid-mechanical phenomenon of self-gravity: Perturbations in the distribution of (dark) matter are amplified by accretion, and this accretion process reflects properties of the cosmological model.

This chapter gives an overview over cosmological models in Section~\ref{bjoern_cosmo} and statistical concepts for describing the distribution of matter in Section~\ref{bjoern_statistics}. Fluid mechanics in its application in discussed in Sections~\ref{bjoern_relativistic},~\ref{bjoern_linear} and~\ref{bjoern_nonlinear}. Open questions are reviewed in Section~\ref{bjoern_questions} before a summary in Section~\ref{bjoern_summary} concludes the chapter.

\section{Cosmology}\label{bjoern_cosmo}
Modern cosmology \citep{1999coph.book.....P, 2000cils.book.....L, 2003euff.book.....B, 2005rgc..book.....C, 2006gere.book.....H, 2008gafo.book.....L} is based on general relativity as the gravitational theory which links the metric on the largest scales to the gravitating fluids. On scales larger than a few hundred $\mathrm{Mpc}/h$ the distribution of matter appears uniform to a freely falling observer who perceives the metric to be homogeneous and isotropic (the Copernican principle), and to vary only with time. The line element $\mathrm{d}s^2 = g_{\mu\nu}\dd x^\mu\dd x^\nu$ which incorporates the Copernican principle is the Robertson-Walker line element, which is most conveniently expressed in spherical coordinates $(t,\chi,\theta,\phi)$ with the comoving distance $\chi$ as the radial coordinate,
\begin{equation}
\mathrm{d}s^2 = 
c^2\dd t^2 - a^2(t)\left(\frac{1}{1-K\chi^2}\dd \chi^2 + r^2\dd\theta^2 + r^2\sin^2\theta\dd\varphi^2\right),
\end{equation}
with the scale factor $a(t)$, which describes the scaling of all physical distances with cosmic time $t$ that every freely falling observer reads off from her or his clock. By convention, the scale factor assumes a value of $a=1$ today and the relation to redshift is given by $a = 1/(1+z)$.

The gravitational field equation links the local curvature encoded in Ricci-tensor $R_{\mu\nu}$ and the Ricci-scalar $R = g^{\mu\nu}R_{\mu\nu}$ to the energy-momentum tensor $T_{\mu\nu}$ of the cosmological fluids,
\begin{equation}
G_{\mu\nu} = \frac{8\pi G}{c^4}\: T_{\mu\nu}
\quad\mathrm{with}\quad
G_{\mu\nu} = R_{\mu\nu} - \frac{R}{2}g_{\mu\nu},
\end{equation}
while obeying local energy-momentum conservation $\nabla^\mu T_{\mu\nu}=0$ expressed through a covariant divergence. The source $T_{\mu\nu}$ of the gravitational field is assumed to be an ideal, relativistic fluid which is parameterised by density and pressure,
\begin{equation}
T_{\mu\nu} = \left(\rho c^2 + p\right)\upsilon_\mu\upsilon_\nu - g_{\mu\nu} p
\end{equation}
and depends on the 4-velocity $\upsilon_\mu$ which is simply $(c,\vec{0})$ for a comoving observer. An ideal fluid which is covariantly conserved, $\nabla^\mu T_{\mu\nu}=0$, follows naturally from a Lagrange density $\mathcal{L} = \mathcal{L}(\varphi,\nabla^\mu\varphi)$ that depends only on the fields $\varphi$ and their derivatives $\nabla^\mu\varphi$, but not explicitly on the coordinates $x^\mu$.

Substitution of the Robertson-Walker-metric into the field equation yields the Friedmann equations, firstly,
\begin{equation}
\left(\frac{\dot{a}}{a}\right)^2 = \frac{8\pi G}{3}\rho -\frac{K}{a^2},
\end{equation}
as an expression for the time-derivative of the logarithmic scale factor, i.e. the Hubble function $H(a)=\dot{a}/a$. The second Friedmann-equation expresses the deceleration parameter $q(a)=-\ddot{a}a/\dot{a}^2$ in terms of the trace of the energy-momentum tensor,
\begin{equation}
\frac{\ddot{a}}{a} = -\frac{4\pi G}{3} (\rho + 3p).
\end{equation}
These two relations constitute the Friedmann-Lema{\^i}tre cosmologies that are based on the spherically symmetric, spatially homogeneous metric as a solution to the field equation of general relativity, with homogeneous ideal fluids, whose density $\rho$ and pressure $p$ contribute to the energy-momentum tensor $T_{\mu\nu}$. It is worth noting that they are among the simplest solutions of general relativity and constitute a system with pure Ricci-curvature.

Relativistic local energy-momentum conservation $\nabla^\mu T_{\mu\nu}=0$ can be recovered from the Friedmann-equations, because the gravitational field equation is constructed to respect it, either in the form of the adiabatic equation,
\begin{equation}
\frac{\dd}{\dd a}\left(a^3\rho(a)\right) + p \frac{\dd}{\dd a}\left(a^3\right) = 0
\end{equation}
or, equivalently, in the shape of a continuity equation,
\begin{equation}
\dot{\rho} + 3H(a)\left(p+\rho\right) = 0,
\end{equation}
under the assumption of ideal, relativistic fluids and the FLRW-symmetries.

Empirically, the relation between pressure and density is parametrised by the equation of state parameter $w$:
\begin{equation}
p = w\:\rho c^2,
\end{equation}
because this relation is the only free choice under the FLRW-symmetry assumptions. Pressureless dark matter is characterised by $w=0$ and the value $w\equiv -1$ corresponds to the cosmological constant $\Lambda$. For the curvature $K$ to vanish, the density $\rho$ has to be equal to the critical density $\rho_\mathrm{crit}(a)=3H^2_0/(8\pi G)\simeq1.8784\times10^{-26}~\mathrm{kg}/\mathrm{m}^3 \simeq2.7745\times10^{11}h^2~M_\odot/\mathrm{Mpc}^3$, which implies the scaling of the density parameter $\Omega_w(a)=\rho(a)/\rho_\mathrm{crit}(a)$ of each fluid,
\begin{equation}
\frac{\Omega_w(a)}{\Omega_w} = \frac{H_0^2}{a^{3(1+w)}H^2(a)},
\label{eqn_adiabatic}
\end{equation}

The relation between equation of state $w$ and deceleration parameter $q$ is given by
\begin{equation}
2(1 + q) = 3(1 + w),
\end{equation}
if the density of the fluid corresponds to the critical density. Fluids with $w<-1/3$ are interesting as an explanation for the accelerated expansion of the universe observed today, as $\ddot{a}>0$ and $q<0$, and the cosmological constant in particular is described by a fluid $w=-1$, although it is formally already included as a term $\Lambda g_{\mu\nu}$ in the field equation. 

Curvature can be formally described by including a fluid with density $\Omega_K=1-\Omega_w$ and assigning it an equation of state of $-1/3$, although curvature is not a physical substance and although $\Omega_K$ can be, in contrast to all other density parameters negative. In addition, Nature does not use this particular degree of freedom, as cosmic inflation has driven the parameter $\Omega_K$ to very small values, as current limits on $\Omega_K$ are well below the percent level.

Table~\ref{table_eos} summarises the most important cosmological fluids, their equation of state, their dependence on scale factor in the Hubble function as well as their deceleration parameter.

\begin{table}
\begin{center}
\begin{tabular}{lllcc}
\hline\hline
fluid		& $\rho(a)$	&	$H(a)$		 		& $w$ 	& $q$ \\
\hline
radiation	& $\propto a^{-4}$	&	$\propto a^{-2}$		& $+1/3$& $1$ \\
matter		& $\propto a^{-3}$	&	$\propto a^{-3/2}$		& $0$	& $1/2$ \\
curvature	& $\propto a^{-2}$	&	$\propto a^{-1}$		& $-1/3$& $0$ \\
dark energy	& $\propto a^{-2\ldots0}$	&	$\propto a^{-1\ldots0}$	& $-1/3\ldots -1$	& $0\ldots -1$ \\
$\Lambda$	& $=\mathrm{const}$	&	$=\mathrm{const}$		& $-1$	& $-1$ \\
phantom energy	&	$\propto a^{q}, q>0$	& grows & $< -1$& $<-1$ \\
\hline
\end{tabular}
\end{center}
\caption{Summary of the relevant cosmological fluids, the scaling of their density $\rho(a)$, their influence on the Hubble function $H(a)$, their equation of state $w$ and the resulting deceleration parameter $q$.}
\label{table_eos}
\end{table}

\subsection{Dark energy and its parametrisation}
Using the results from the last section, the Hubble function $H(a) = \dd\ln a/\dd t$ for a universe filled with radiation, pressureless dark matter and dark energy with a time-evolving equation of state $w(a)$ can be constructed to be
\begin{equation}
\frac{H^2(a)}{H_0^2} = 
\frac{\Omega_\gamma}{a^{4}} + 
\frac{\Omega_m}{a^{3}} + 
\Omega_\varphi\exp\left(3\int_a^1\dd\ln a\:\left[1+w(a)\right]\right) \stackrel{w(a)=w_0}{=}
\frac{\Omega_\gamma}{a^4} + \frac{\Omega_m}{a^3} + \frac{\Omega_\varphi}{a^{3(1+w_0)}},
\end{equation}
in spatially flat dark energy cosmology with the the matter density $\Omega_m$ and the dark energy density $\Omega_\varphi = 1 - \Omega_\gamma - \Omega_m$. Today's estimates $\Omega_m = 0.25$ and $\Omega_\varphi = 0.75$ allow to quantify the time of matter-dark energy equality $a_{m\varphi} \simeq 0.7$ (for $w=-1$), defined by $\Omega_m(a_{m\varphi}) = \Omega_\varphi(a_{m\varphi})$, while the radiation density is small, $\Omega_\gamma\simeq10^{-5}$.

The relation between comoving distance $\chi$ (given in terms of the Hubble distance $\chi_H=c/H_0=2.9969~\mathrm{Gpc}/h$) and scale factor $a$ is then given by
\begin{equation}
\chi = c\int_a^1\:\frac{\dd a}{{a}^2 H(a)},
\end{equation}
with the speed of light $c$. The dark energy equation of state $w(a)$ is conveniently approximated by its first order Taylor expansion with respect to the scale-factor $a$,
\begin{equation}
w(a) = w_0 + (1-a) w_a,
\end{equation}
introduced by \citet{2001IJMPD..10..213C} and \citet{2003MNRAS.346..573L}, for non-interacting dark energy models with a slow time evolution $w(a)$. Incidentially, this parameterisation allows a complete integration of the Hubble-function $H(a)$. The conformal time $\eta$ in units of the Hubble time $t_H=1/H_0$ follows in analogy to the comoving distance $\chi$ from the differential $\dd\eta=\dd t/a=\dd a/(a^2H(a))$,
\begin{equation}
\eta = \int_a^1\: \frac{\dd a}{a^2 H(a)},
\end{equation}
such that $\chi=c\eta$ and one recovers Minkowskian light propagation in these coordinates.

\begin{figure}
\begin{center}
\resizebox{0.9\hsize}{!}{\includegraphics{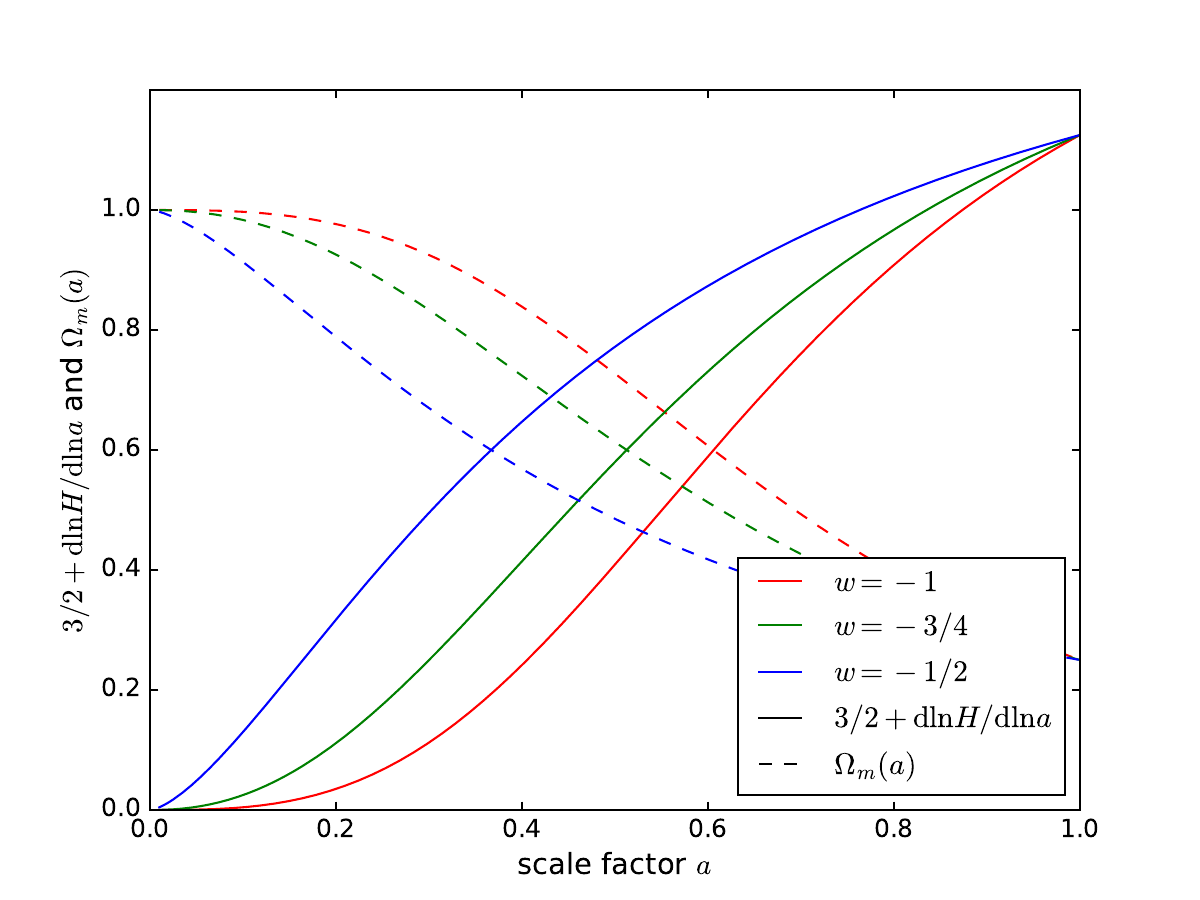}}
\end{center}
\caption{The logarithmic derivative $3/2+\dd\ln H/\dd\ln a$ (solid lines) and the matter density parameter $\Omega_m(a)$ (dashed lines), for a cosmology with a cosmological constant $\Lambda$ with $w_0=-1$ (red lines), dark energy with $w_0=-3/4$ (green lines) and with $w_0=-1/2$ (blue lines).}
\label{fig_hubble}
\end{figure}

Fig.~\ref{fig_hubble} shows the logarithmic derivative $\dd\ln\tilde{H}/\dd\ln a=3/2+\dd\ln H/\dd\ln a$ of the Hubble function, where the added term $3/2$ removes the rapid scaling $H\propto a^{-3/2}$ in the matter dominated epoch, as well as the matter density parameter $\Omega_m(a)$ as a function of time and for a range of dark energy equation of state parameters.

\section{Statistical description of structures}\label{bjoern_statistics}
The fluctuations of the cosmic density field $\delta(\mathbf{x})$, for which one chooses the relative deviation of the density field $\rho(\mathbf{x})$ from the mean background density $\bra\rho\ket=\Omega_m\rho_\mathrm{crit}$,
\begin{equation}
\delta(\mathbf{x}) = \frac{\rho(\mathbf{x})-\bra\rho\ket}{\bra\rho\ket},
\end{equation}
are assumed to be Gaussian with a certain correlation length, meaning that the probability of finding the amplitudes $\delta(\mathbf{x}_1)$ and $\delta(\mathbf{x}_2)$ and positions $\mathbf{x}_1$ and $\mathbf{x}_2$ in a hypothetical ensemble of universes is given by a bivariate Gaussian probability density,
\begin{equation}
p(\delta(\mathbf{x_1}),\delta(\mathbf{x_2})) =
\frac{1}{\sqrt{(2\pi)^2\mathrm{det}(Q)}}
\exp\left[
-\frac{1}{2}
\left(\begin{array}{c}\delta(\mathbf{x}_1) \\ \delta(\mathbf{x}_2)\end{array}\right)^t
C^{-1}
\left(\begin{array}{c}\delta(\mathbf{x}_1) \\ \delta(\mathbf{x}_2)\end{array}\right)
\right]
\end{equation}
with the covariance matrix $C$:
\begin{equation}
C = \left(
\begin{array}{cc}
\bra\delta(\mathbf{x}_1)^2\ket & \bra\delta(\mathbf{x}_1)\delta(\mathbf{x}_2)\ket \\
\bra\delta(\mathbf{x}_2)\delta(\mathbf{x}_1)\ket & \bra\delta(\mathbf{x}_2)^2\ket
\end{array}
\right)
\end{equation}
The off-diagonal variance in $C$ is the correlation function $\xi(\mathbf{x}_1,\mathbf{x}_2)\equiv\bra\delta(\mathbf{x}_1)\delta(\mathbf{x}_2)\ket$ of the random field, which describes how fast with increasing distance $\left|\mathbf{x}_2-\mathbf{x}_1\right|$ the field loses its memory on the amplitude at $\mathbf{x}_1$. A length scale in $\xi(\mathbf{x}_1,\mathbf{x}_2)$ can be interpreted as a correlation length. Due to the Cauchy-Schwarz inequality,
\begin{equation}
\bra\delta(\mathbf{x}_1)\delta(\mathbf{x}_2)\ket^2
\leq
\bra\delta(\mathbf{x}_1)^2\ket\bra\delta(\mathbf{x}_2)^2\ket
\quad\rightarrow\quad
r
= 
\frac{\bra\delta(\mathbf{x}_1)\delta(\mathbf{x}_2)\ket}
{\sqrt{\bra\delta(\mathbf{x}_1)^2\ket\bra\delta(\mathbf{x}_2)^2\ket}} 
\end{equation}
the correlation function is always smaller than the geometrical mean of the variances at a single point, i.e. the covariance $C$ is positive definite, and the Pearson-correlation coefficient $\left|r\right|$ is smaller than unity. 

Clearly, if the correlation function $\xi(\mathbf{x}_1,\mathbf{x}_2)$ vanishes, the Gaussian probability density separates, $p(\delta(\mathbf{x_1}),\delta(\mathbf{x_2})) = p(\delta(\mathbf{x_1})) p(\delta(\mathbf{x_2}))$ and the amplitudes are mutually uncorrelated and the field amplitudes follow independently from univariate Gaussian distributions.

In homogeneous Gaussian random fields the correlation function $\xi(\mathbf{x}_1,\mathbf{x}_2)$ only depends on the separation $\mathbf{r} = \mathbf{x}_2 -\mathbf{x}_2$ between the two points and not the location. Then, the two variances $\bra\delta(\mathbf{x}_1)^2\ket$ and $\bra\delta(\mathbf{x}_2)^2\ket$ are identical. If in addition the random field has isotropic fluctuation properties, the correlation function only depends on the magnitude of $\mathbf{r}$ and not its direction. In all of the above, the averaging brackets $\bra\ldots\ket$ denote the average over realisations of the density field. 

The knowledge of the variance is sufficient because all moments of a Gaussian distributed random variable with zero mean are proportional to the variance, $\bra\delta^{2n}\ket\propto\bra\delta^2\ket^n$. Hence the characteristic function $\varphi(t)=\int\dd\delta p(\delta)\exp(\ci t\delta)=\sum_n \bra\delta^n\ket (\ci t)/n!$ only requires the estimation of the variance $\bra\delta^2\ket$ for reconstructing $p(\delta)\dd\delta$ from the moments $\bra\delta^{2n}\ket$ by inverse Fourier transform.

Homogeneous random fields possess the interesting property of mutually uncorrelated Fourier modes. Applying a Fourier-transform to the density contrast $\delta(\mathbf{x})$,
\begin{equation}
\delta(\mathbf{k}) = \int\dd^3x\:\delta(\mathbf{x})\exp(-\ci\mathbf{kx})
\quad\leftrightarrow\quad
\delta(\mathbf{x}) = \int\frac{\dd^3 k}{(2\pi)^3}\:\delta(\mathbf{k})\exp(+\ci\mathbf{kx}),
\end{equation}
and computing the variance between two Fourier modes $\delta(\veck_1)$ and $\delta(\veck_2)$
\begin{equation}
\bra\delta(\veck_1)\delta^*(\veck_2)\ket=(2\pi)^3\delta_D(\veck_1-\veck_2)P(\veck_1),
\label{eqn_pspectrum}
\end{equation} 
shows that it is proportional to the Dirac $\delta$-function, indicating their statistical independence. The proportionality is given by the spectrum,
\begin{equation}
P(\veck) = \int\dd^3r\: \xi(\mathbf{r})\exp(-\ci\mathbf{kr}),
\end{equation}
which follows as the Fourier-transform of the correlation function $\xi(\mathbf{r})$. If in addition the random field is isotropic, $P(k)$ only depends only the wave number $k$ instead of the wave vector $\veck$. In this case, the angular integrations in eqn.~(\ref{eqn_pspectrum}) can be carried out by introducing spherical coordinates in Fourier-space, yielding:
\begin{equation}
P(k) = 2\pi\int_{0}^\infty r^2\dd r\: \xi(r) j_0(kr),
\end{equation}
with the spherical Bessel function of the first kind $j_0(kr)$ of order $\ell=0$, being equal to $\mathrm{sinc}(kr)=\sin(kr)/(kr)$ \citep{1972hmf..book.....A}.

Cosmological inflation provides a mechanism for generating Gaussian fluctuation fields with the spectrum $P(k)$,
\begin{equation}
P(k)\propto k^{n_s}T^2(k)
\end{equation}
\citep{2008arXiv0810.3022B, 2010arXiv1001.5259L}, with the CDM transfer function $T(k)$. $T(k)$ describes the scale-dependent suppression of the growth of small-scale modes between horizon-entry and matter-radiation equality by the Meszaros-mechanism. It is well approximated with the polynomial fit proposed by \citet{1986ApJ...304...15B},
\begin{displaymath}
T(q) = \frac{\ln(1+2.34q)}{2.34q}\left(1+3.89q+(16.1q)^2+(5.46q)^3+(6.71q)^4\right)^{-\frac{1}{4}},
\label{eqn_cdm_transfer}
\end{displaymath}
or the more accurate fit described by \citet{1998ApJ...496..605E, 1999ApJ...511....5E} incorporating baryonic wiggles for flat cosmological models with low matter density $\Omega_m$. The asymptotic behaviour of the transfer function is such that $T(k)\propto\mathrm{const}$ for $k\ll 1$ and $T(k)\propto k^{-2}$ at $k\gg 1$, such that $P(k)\propto k^{n_s}$ on large scales and $P(k)\propto k^{n_s-4}$ on small scales. In particular the scale-free behaviour of $P(k)$ on small scales is a feature of cold dark matter models, because any thermal motion would wipe out structures on small scales and would lead to an exponential cut-off in the spectrum.

The wave vector is rescaled with the shape parameter $\Gamma\simeq\Omega_m h$, which corresponds to the horizon size at the time of matter-radiation equality $a_{\gamma m}$, and describes the peak shape of the CDM power spectrum $P(k)$. The corrections of a non-negligible baryon density $\Omega_b$ on the peak shape are described by \citet{1995ApJS..100..281S},
\begin{equation}
\Gamma=\Omega_m h\:\exp\left[-\Omega_b\left(1+\frac{\sqrt{2h}}{\Omega_m}\right)\right],
\end{equation}
where $\Gamma$ is measured in units of $(\mathrm{Mpc}/h)^{-1}$, such that $q=k/\Gamma$ is a dimensionless wave vector. The spectrum is usually normalised to the variance of the linearly evolved density field at zero redshift on a scale of $R=8~\mathrm{Mpc}/h$,
\begin{equation}
\sigma^2_R = \frac{1}{2\pi^2}\int_0^\infty\dd k\:k^2 P(k) W^2(kR),
\end{equation}
with a Fourier transformed spherical top hat filter function, $W(x)=3j_1(x)/x$. $j_1(x)$ is the spherical Bessel function of the first kind of order $\ell=1$  \citep{1972hmf..book.....A, 2005mmp..book.....A}. This particular definition of $\sigma(R)$, along with the fact that the power spectrum has the dimension of a volume, motivates the definition of the dimensionless power spectrum $\Delta^2(k)\propto k^3P(k)$,
\begin{equation}
\Delta^2(k) = \frac{k^3}{2\pi^2}P(k)
\quad\rightarrow\quad
\sigma^2_R = \int_0^\infty\dd\ln k\:\Delta^2(k) W^2(kR),
\end{equation}
such that $\Delta^2(k)$ reflects the fluctuation variance per logarithmic band in $k$, $\dd\sigma_R^2/\dd\ln k \propto \Delta^2$. As the spectrum $P(k)$ has the unit of a volume and is conventionally given in units of $(\mathrm{Mpc}/h)^3$, the variance $\Delta^2(k)$ is dimensionless.

\section{Relativistic fluid mechanics and the nonrelativistic limit}\label{bjoern_relativistic}
The equations of relativistic fluid mechanics results from relativistic energy momentum conservation $\nabla^\mu T_{\mu\nu} = 0$ by projecting this relation onto the 4-velocity $\upsilon_\mu$ of an observer (which in our case is a FLRW-observer) and perpendicular to it. Specifically, $\upsilon^\nu\nabla^\mu T_{\mu\nu}=0$ yields the continuity equation and $(\upsilon^\alpha\upsilon^\nu/c^2 - g^{\alpha\nu})\nabla^\mu T_{\mu\nu}=0$ the Euler-equation. 

If the perturbations of the gravitational field are weak, $\Phi\ll c^2$, the metric will only contain terms involving the Newtonian gravitational potential $\Phi$ which enters the Euler-equation through a gradient $\nabla\Phi$ as the remnant of the covariant derivative $\nabla^\mu$, and the error by using and FLRW-observer's 4-velocity is small, implying in particular that one can use cosmic time instead of the proper time if in addition the fluid velocities are small compared to $c$. The projections will only use in this limit the covariant derivative $\nabla^\mu$ defined by the FLRW-metric and ignore perturbations in the metric, reducing the equations of fluid mechanics by Newtonian relations on an expanding background, with the cosmic time as the global time variable.

In the limit of small metric perturbations the gravitational field equation will be replaced by $\Box\Phi = 4\pi G\rho$, with only the matter density being active as the source of the gravitational field, while pressure is suppressed by a factor of $1/c^2$ relative to the matter density. Furthermore, since most of the gravitating matter is nonrelativistic, pressure does not play a role as a source of gravitational perturbations. If in addition one assumes that retardation effects in the propagation of the gravitational field is not important on small scales, one recovers the classical Poisson-equation $\Delta\Phi = 4\pi G\rho$ as the gravitational field equation. The retardation scale in this is set by the Hubble scale $c/H_0$.

Specifically, the Poisson-equation reformulated in terms of comoving distances reads for the dimensionless potential $\varphi = \Phi/c^2$
\begin{equation}
\Delta\varphi = 
\frac{3}{2}\Omega_m(\eta)\:\frac{(aH)^2}{c^2}\delta = 
\frac{3\Omega_m}{2\chi_H^2}\frac{\delta}{a},
\end{equation}
where the strength of the gravitational potential is influence by the cosmology through the evolution of the ambient matter density and the scale is set by the Hubble distance $\chi_H = c/H_0$. Newton's constant $G$ has been replaced with the definition of the critical density, and absorbed into the ambient density into $\Omega_m(\eta)$ with the conformal time $\eta$ as the time variable. The replacement of $\Omega_m(\eta)$ by $\Omega_m$ uses the adiabatic evolution of the matter density in eqn.~\ref{eqn_adiabatic}.

\section{Linear structure formation}\label{bjoern_linear}
\subsection{Linearised structure formation equations}
The formation of cosmic structures is based on gravitational amplification of seed perturbations in the cosmic density field. Following \citet{2002PhR...367....1B}, the most convenient choice of variables is to use the physical density field, the velocity fields and the potential on one side, the conformal time as the time variable and the comoving distance for the differential operators on the other. Being a hydrodynamical self-gravitating phenomenon, structure formation is described in the comoving frame by the system of differential equations composed of $(i)$ the continuity equation
\begin{equation}
\frac{\partial}{\partial\eta}\delta + \mathrm{div}\left[(1+\delta)\vecv\right] = 0,
\label{eqn_continuity}
\end{equation}
which relates the time-evolution of the density field to the divergence of the matter fluxes $(1+\delta)\vecv$, $(ii)$ the Euler-equation
\begin{equation}
\frac{\partial}{\partial\eta}\vecv + aH\vecv + (\vecv\nabla)\vecv = -\nabla\Phi,
\label{eqn_euler}
\end{equation}
which describes the evolution of the peculiar velocity field $\vecv$ from the gradient $\nabla\Phi$ of the peculiar gravitational potential $\Phi$, acting on a fluid element, and finally $(iii)$ the comoving Poisson-equation
\begin{equation}
\Delta\Phi = \frac{3H_0^2\Omega_m}{2a}\delta,
\end{equation} 
which gives the gravitational potential $\Phi$ induced by the matter distribution $\delta$. The three equations are sufficient to describe the dynamics of the three relevant fields $\delta$, $\vecv$ and $\Phi$, because there are no dissipative and pressure forces due to the collisionlessness of dark matter, and it is not necessary to track the energy balance or to introduce and an equation of state parametrising the pressure-density relation. The expanding background is described by the Hubble function $H(a)$ and the usage of Newtonian dynamics and Newtonian gravity is well justified due to the small velocities $\upsilon\ll c$ and the smallness of gravitational potentials $\Phi\ll c^2$ on sub-horizon scales $k\geq 2\pi/\chi_H$, where  retardation effects do not play a role.

Linearisation of the structure formation equations by substituting a perturbative expansion of the density- and velocity field yields the linearised continuity equation,
\begin{equation}
\frac{\partial}{\partial\eta}\delta + \mathrm{div}\vecv = 0,
\end{equation}
and the linearised Euler-equation,
\begin{equation}
\frac{\partial}{\partial\eta}\vecv + aH\vecv = -\nabla\Phi,
\end{equation}
which are valid if the overdensity $\delta$ is small, $\delta\ll 1$. The Newtonian Poisson-equation is always linear, or the superposition principle would not hold.

\subsection{Growth equation and its solution}
The divergence of the linearised Euler-equation can be combined with the linearised continuity equation and its time derivative as well as the Poisson equation to yield the growth equation as a second-order ordinary differential equation. As spatial derivatives cancel, the solution describes the homogeneous growth $\delta(\mathbf{x},a)=D_+(a)\delta(\mathbf{x},a=1)$ of the density field in the linear regime of structure formation, $\left|\delta\right|\ll 1$ \citep{1980PhyS...21..720P, 1997PhRvD..56.4439T, 2003MNRAS.346..573L},
\begin{equation}
\frac{\dd^2}{\dd a^2}D_+(a) + \frac{1}{a}\left(3+\frac{\dd\ln H}{\dd\ln a}\right)\frac{\dd}{\dd a}D_+(a) = 
\frac{3}{2a^2}\Omega_m(a) D_+(a).
\label{eqn_growth}
\end{equation}

The growing solution $D_+(a)$ of the growth equation's two solutions is referred to as the growth function and assumes the simple solution $D_+(a)=a$ in the SCDM cosmology, where $\Omega_m\equiv 1$ and $H(a)/H_0 = a^{-3/2}$. This scaling motivates the usage of the scale factor $a$ as the preferred time variable, and suggests to transform the differentials with $\dd/\dd\eta = a^2H(a)\dd/\dd a$. For that reason, the initial conditions for solving the growth equation are given by $D_+(0) = 0$ and $\dd D_+(0)/\dd a = 1$, due to the domination of the $\Omega_m$-term in $H(a)$ at early times. Is is a peculiar result that, due to the additional factor $1/a$ in the Poisson-equation, gravitational potentials are constant during linear structure formation in cosmologies with $\Omega_m = 1$. By convention, the growth function is normalised to the value $D_+=1$ today, $a=1$.

$\Omega_m(a)$ on the right hand side acts as a driving term, because gravitational potentials are strong if $\Omega_m(a)$ is large, making it easier for the cosmic large-scale structure to form. The term $3+\dd\ln H/\dd\ln a$ is sometimes referred to as Hubble-drag, which suppresses the formation of structures if their dynamical time scale is larger than the time scale of the Hubble expansion. It is worth emphasising, however, that only a change in the expansion rate of the Universe affects structure formation: Expansion at a constant Hubble-rate would not affect the contrast $\delta$ of structures at all.

The second solution $D_-(a)$ scales $\propto 1/a$ in the matter dominated phase of the Universe and does not play a role at late times due to its fast decay.

It is worth noting that the Hubble function $H(a)$ and the growth function $D_+(a)$ convey the same information, as one can compute one from the other. In fact, \citet{1980lssu.book.....P} shows that $D_+(a)$ is given by
\begin{equation}
D_+(a) \propto H(a)\int_a^1\frac{\dd a}{\left(aH(a)\right)^3}.
\end{equation}
This is due to the fact that in standard cosmologies the dependence $\Omega_m(a)$ on time is given entirely by the Hubble function $H(a)$, and that in Newtonian gravity the same term determines the term on the right hand side driving structure formation. This ceases to be applicable, however, in modified gravity theories, or in cosmologies with interactions between the cosmological fluids. Naturally, a signature of these cosmologies would be a mismatch between the information conveyed by $H(a)$ and $D_+(a)$.

Homogeneous structure formation corresponds to independently growing Fourier modes,
\begin{equation}
\delta(\mathbf{x},a) = D_+(a)\delta(\mathbf{x},a=1) \longrightarrow\delta(\veck,a) = D_+(a)\delta(\veck,a=1),
\label{eqn_homo_growth}
\end{equation}
which conserves every statistical property of the initial conditions, in particular Gaussianity. The Gaussianity of the initial density perturbations is a consequence of inflation, where a large number of uncorrelated quantum fluctuations are superimposed, yielding a Gaussian amplitude distribution due to the central limit theorem. In fact, homogeneous growth in the linear regime is the reason why investigation of inflationary processes in structure is possible by observing the large-scale structure today, even after the cosmic time $1/H_0$ has passed.

\begin{figure}
\begin{center}
\resizebox{0.9\hsize}{!}{\includegraphics{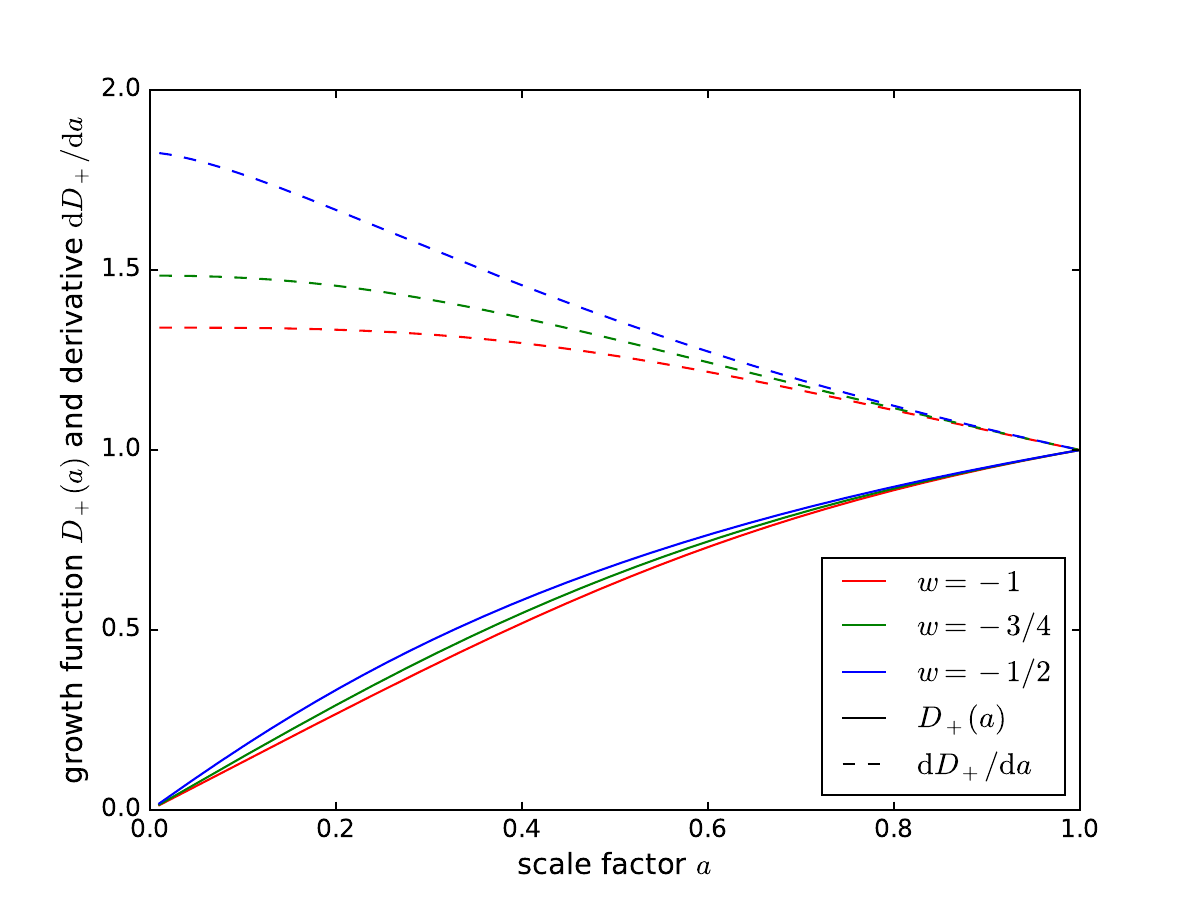}}
\end{center}
\caption{Linear growth function $D_+(a)$ (solid lines) and the evolution $D_+/a$ of potentials (dashed lines), for varying equation of state parameter $w$ of the dark energy fluid: cosmological constant $\Lambda$ with $w_0=-1$ (red lines), dark energy with $w_0=-3/4$ (green lines) and with $w_0=-1/2$ (blue lines).}
\label{fig_growth}
\end{figure}

Fig.~\ref{fig_growth} shows the growth function $D_+(a)$ in comparison to the derivative $D_+/a$ for cosmologies with $\Omega_m=0.25$ and different values of the equation of state $w$ of the dark energy fluid.

A convenient way for approximating the growth function is the $\gamma$-parameter, introduced by \citet{1980lssu.book.....P} in the study of peculiar velocities:
\begin{equation}
\frac{\dd\ln D_+}{\dd\ln a} \simeq \Omega_m(a)^\gamma,
\end{equation}
with $\gamma\simeq0.6$ in $\Lambda$CDM. Solving this equation for the growth function yields
\begin{equation}
D_+(a) = \exp\left(\int_0^a\dd\ln a\: \Omega_m(a)^\gamma\right).
\end{equation}
In dynamic dark energy models, $\gamma$ can be approximated by $\gamma=0.55+0.05(1+w(z=1))$ with the dark energy equation of state parameter taken at unit redshift \citep{2005PhRvD..72d3529L}.

The effect of adding a fluid with a negative equation of state is a slower growth in the recent cosmic past and a faster growth in the remote past (if the growth function is normalised to unity today). Fig.~\ref{fig_growth} shows the growth function $D_+(a)$ and the evolution of the potentials $D_+/a$. This dependence on the cosmological model is generated by the scaling of $\Omega_m$ with time and the magnitude of the term $3+\dd\ln H/\dd\ln a$, which can be shown to be equation to $2-q$ with the deceleration parameter $q$. Physically, they originate from the comoving Poisson equation and the cosmology-dependence of the relativistic Euler-equation: A higher matter density generates stronger gravitational fields for a given perturbation in $\delta$, thus enhancing structure formation. An accelerating universe makes it difficult for structures to form, and ultimately cosmic structure formation is truncated by the dark energy-domination of the cosmic expansion. Conversely, the early decelerating phase of the Universe during matter domination made it easy for structures to form.

\subsection{Peculiar velocities}
Matter fluxes in the large-scale structure drive structure formation: If they converge, they transport matter into a volume and increase the local density, according to the continuity equation. In order to investigate the properties of matter fluxes one can carry out a Helmholtz-decomposition of the velocity field into its curl and gradient components $\theta = \mathrm{div}\vecv$ and $\vecw = \mathrm{rot}\vecv$. From the Euler-equation one obtains the evolution equation for the divergence of the matter fluxes \citep{2002PhR...367....1B},
\begin{equation}
\frac{\partial}{\partial\eta}\theta + aH\theta + \frac{3H_0^2\Omega_m}{2a} \delta =0
\end{equation}
and the corresponding equation for the vorticity $\vecw$,
\begin{equation}
\frac{\partial}{\partial\eta}\vecw + aH\vecw = 0.
\end{equation}
With the definition of the differential of the conformal time, $\dd a = a^2H\dd\eta$, one immediately notices that $\dd\ln\vecw=-\dd\ln\ a$, and hence $\vecw\propto 1/a$ during matter domination: Vorticity can not be generated in linear structure formation in collisionless fluids, and the flows are necessarily laminar. The divergence $\theta$ can be linked to the evolution of the density field using the continuity equation,
\begin{equation}
\theta = -a H \frac{\dd\ln D_+}{\dd\ln a}\delta,
\label{eqn_velocity_divergence}
\end{equation}
which underlines the fact that in the linear regime of structure formation, the velocity field is the gradient of a potential, and the velocity divergence is proportional to the density contrast.

\section{Nonlinear structure formation}\label{bjoern_nonlinear}

\subsection{Phenomenology of nonlinear structure formations}
As long the structure formation is linear, the growth is homogeneous and conserves the Gaussianity of the initial conditions. Nonlinear structure formation implies inhomogeneous growth and the emergence of non-Gaussian features, which is illustrated by a number of arguments: Non-linearity implies inhomogeneity, because if e.g. a void reaches underdensities close to $\delta\simeq-1$ (corresponding to $\rho\simeq 0$), the linearisation fails and the growth has to slow down locally. Inhomogeneity implies non-Gaussianity because the initially Gaussian distribution $p(\delta)\dd\delta$ becomes wider with increasing amplitudes $\delta$, but the density $\delta$ can not be more negative than $-1$, requiring the amplitude distribution $p(\delta)\dd\delta$ to become asymmetric and to acquire a nonzero skewness. For completing the argument one immediately notices that in inhomogeneous growth, i.e. a position dependence of the growth rate $D_+(\vecx,a)$, the Fourier-modes $\delta(\veck,a)$ become coupled, violating the central limit theorem such that the superposition of Fourier-modes yields a non-Gaussian amplitude distribution.

\subsection{Eulerian perturbation theory}
The non-linearities in the continuity and Euler-equation make a closed analytical solution impossible. It is possible, however, to obtain a perturbative solution \citep{1987JMP....28.2714B, 1992ApJ...394L...5B, 1995PhR...262....1S, 1995MNRAS.276...39C, 1997GReGr..29..733E, 2002PhR...367....1B} to the structure formation equations, which contains the mode coupling mechanism and describes the generation of non-Gaussianities in nonlinear structure formation. Starting point is a perturbative expansion of the type
\begin{equation}
\delta(\vecx,a) = 
\sum_{n=1}^\infty \delta^{(n)}(\vecx,a) \simeq 
\sum_{n=1}^\infty D_+^n(a)\delta^{(n)}(\vecx)
\label{eqn_perturbation_series}
\end{equation}
where the last step holds exactly in universes with $\Omega_m=1$ and is approximately valid in dark energy cosmologies. $\delta^{(n)}(\vecx)$ is proportional to the $n$th power $\delta(\vecx)^n$ of the initial conditions. The non-linearities in the continuity- and the Euler-equation translate to convolutions of the density and the velocity fields in Fourier space which couple the individual Fourier modes, violating the central limit theorem and therefore violating Gaussianity. It is worth noting that in the perturbative expansion each field $\delta^{(n)}$ grows homogeneously at the rate $D_+^n(a)$, but the sum does not.

Substituting a perturbation series of the type of eqn.~(\ref{eqn_perturbation_series}) for the density and velocity fields into the fully nonlinear, Fourier-transformed structure formation equations, and sorting the according to the exponent $n$ yields:
\begin{equation}
\delta^{(n)}(\veck) = \int\dd^3q_1\ldots\int\dd^3q_n\:
\delta_D\left(\veck-\sum_{i=1}^n\vecq_i\right)
F_n(\vecq_1,\ldots,\vecq_n)\prod_{i=1}^n\delta(\vecq_i)
\end{equation}
with the mode coupling function $F_n(\vecq_1,\ldots,\vecq_n)$, for which a recursion relation can be obtained. The lowest order symmetrised solutions for $F_n$ are $F_1=1$ and
\begin{equation}
F_2(\vecq_1,\vecq_2) = \frac{5}{7}+\frac{x}{2}\left(\frac{q_1}{q_2}+\frac{q_2}{q_1}\right) + \frac{2}{7}x^2
\quad\mathrm{with}\quad
x = \frac{\vecq_1\cdot\vecq_2}{q_1q_2}
\end{equation}
being the cosine of the angle between $\vecq_1$ and $\vecq_2$. Assuming $q_1=q_2$ for simplicity, the mode coupling function $F_2$ attains the largest value of $F_2=2$ if the wave vectors are parallel ($x=+1$), an intermediate value of $F_2=5/7$ if $\vecq_1\perp\vecq_2$ ($x=0$) and the smallest value of $F_2=0$ if the the wave vectors are antiparallel ($x=-1$). Varying the wave numbers at fixed separation angle $x$ shows that $F_2$ is smallest if $q_1=q_2$, and that the mode coupling increases if the wave numbers are chosen differently. From this point of view, mode-coupling bears resemblance to a resonance phenomenon, where modes with identical direction of propagation experience the strongest coupling. The perturbative solution to the system of equations eqns.~(\ref{eqn_continuity}) and~(\ref{eqn_euler}) in terms of a perturbation series in $\delta$ and $\upsilon$ is possible due to their renormalisation properties, which hold exactly in the case of SCDM ($\Omega_m=1$, $\Omega_\varphi=0$) and approximately for dark energy cosmologies and which are the topic of a number of papers \citep{2006PhRvD..73f3520C, 2006PhRvD..73f3519C, 2008PhRvD..78j3521B}. In these cosmologies, the mode coupling kernels themselves acquire a slow time dependence themselves, which is measured in numerical simulations.

\subsection{Lagrangian perturbation theory}
An alternative way of formulating the perturbative, translinear dynamics of the cosmic density field is Lagrangian perturbation theory, where the central objects are the particle trajectories which link the initial positions $\vecq$ to their positions $\vecx$ at time $\eta$ rather than the density- and velocity fields \citep{1987JMP....28.2714B, 1997GReGr..29..733E, 1989A&A...223....9B, 1994MNRAS.267..811B}. The mapping between the initial position $\vecq$ of a particle and the position $\vecx$ ad time $\eta$ is at lowest order given by
\begin{equation}
\vecx(\vecq,\eta) = \vecq - \nabla\Psi(\vecq,\eta),
\end{equation}
with the displacement potential $\Psi$ which describes the potential flows developing in the large-scale structure in linear structure formation. The lowest order perturbative mapping is referred to as the Zel'dovich-approximation \citep{1970A&A.....5...84Z, 1992MNRAS.254..729B}. The linear solution to $\Psi$ can be derived as $\Delta\Psi(\vecq,\eta) = D_+(\eta)\delta(\vecq)$, using the solution $D_+(\eta)$ to the homogeneous growth equation \citep{1995MNRAS.276..115C}.

Since the equation of motion of a particle in comoving coordinates is given by
\begin{equation}
\frac{\dd^2}{\dd \eta^2}\vecx + aH\frac{\dd}{\dd\eta}\vecx = -\nabla\Phi,
\end{equation}
it is possible to establish a relationship between the displacement potential $\Psi$ and the physical gravitational potential $\Phi$ by taking the divergence of this equation, allowing the replacement of $\Delta\Phi$ with the Poisson equation. Using mass conservation $1+\delta(\vecx,\eta) = 1/J(\vecq,\eta)$ and the Jacobian
\begin{equation}
J(\vecq,\eta)\equiv 
\frac{\dd\vecx}{\dd\vecq} = 
\left(\mathrm{det}\left[\delta_{ij} + \partial_i\partial_j\Psi\right]\right)^{-1},
\end{equation} 
of the mapping between $\vecq$ and $\vecx$ yields the relationship
\begin{equation}
J(\vecq,\eta)\:\mathrm{div}\left[\frac{\dd^2}{\dd\eta^2} + aH\frac{\dd}{\dd\eta}\right]\nabla\Psi = \frac{3}{2}\Omega_m(\eta)(aH)^2(J(\vecq,\eta)-1),
\end{equation}
which illustrates that the trace of the tidal forces $\partial_i\partial_j\Psi$ is responsible for a compression of the cosmic density field. The limit of applicability of Lagrangian perturbation theory is reached when $J=0$, because of the divergence of the density field. This conditions corresponds to the non-invertibility of the mapping $\vecq\rightarrow\vecx$ which occurs when two trajectories cross.

\subsection{non-Gaussian statistics}
In application to statistics, any correlation function of nonlinear fields can reduced to a higher-order correlation function of the linearly evolving fields which obey Gaussian statistics, integrated over momentum space with the mode coupling function as a weighting function. While odd $n$-point correlation functions of Gaussian random fields are equal to zero, even $n$-point functions can be decomposed into products of two-point functions by virtue of the Wick-theorem,
\begin{equation}
\bra\delta(\veck_1)\ldots\delta(\veck_n)\ket = 
\sum_\mathrm{pairs}\:\prod_{i,j\in\mathrm{pairs}}\bra\delta(\veck_i)\delta(\veck_j)\ket
\end{equation}
for which a proof can be found in e.g. \citet{2008cmbg.book.....D} and which constitutes an extension of the well-known relation $\bra \delta^{2n}\ket=(2n-1)!! \bra \delta^2\ket^n$ for the higher moments of a Gaussian random variable $\delta$ with $\bra \delta\ket=0$. In this way it is possible to describe the generation of non-Gaussian statistical properties by substituting the perturbative expansion eqn.~(\ref{eqn_perturbation_series}), which links $n$-point correlation functions or their corresponding polyspectra to higher-order expressions, which render a vanishing odd moment nonzero and add correction terms to the even moments, effectively destroying the Gaussian recursion relation thus making the statistics non-Gaussian.

From a physical point of view the nonlinearities in the continuity- and Euler-equations are responsible for mode-coupling and ultimately, for the generation of non-Gaussian statistics, similarly to the Karman-Hovarth-hierarchy in turbulence theory. The multiplication of the fields $\rho$ and $\vecv$ translates to a convolution in Fourier-space, which links all terms in the perturbative expansion. This effectively causes non-Gaussian statistical property, even for initially Gaussian fields, by the generation of corrections to the moments.

\subsection{Halo formation}
Halos of dark matter form from isolated high-density peaks of the fluctuating matter field, by gravitational collapse under their own gravity. In the limiting case of spherical protohaloes it is possible to solve the collapse equation for an $\Omega_m=1$-cosmology exactly in terms of a parametric solution, and the numerical solution of the collapse equation for dark energy cosmologies shows that there is only a small difference in the collapse overdensity $\delta_c$. Due to the Birkhoff-theorem, only the matter interior of a spherically symmetric perturbation drives the collapse according to an equation of motion of the type
\begin{equation}
\ddot{R} = -\frac{GM}{R^2}.
\end{equation}
Solving this equation in a dimensionless form yields a minimal overdensity for spherical collapse of $\delta_c\simeq1.69$ for $\Omega_m=1$-cosmologies. Consequently, one would consider all regions in the initial field that, under linear growth would reach amplitudes $\delta\lsim\delta_c$ today as collapsed into haloes. The influence of tidal shear fields on the collapse dynamics is minor, but they are responsible for the generation of angular momentum in haloes through the process of tidal torquing.

\begin{figure}
\begin{center}
\resizebox{0.9\hsize}{!}{\includegraphics{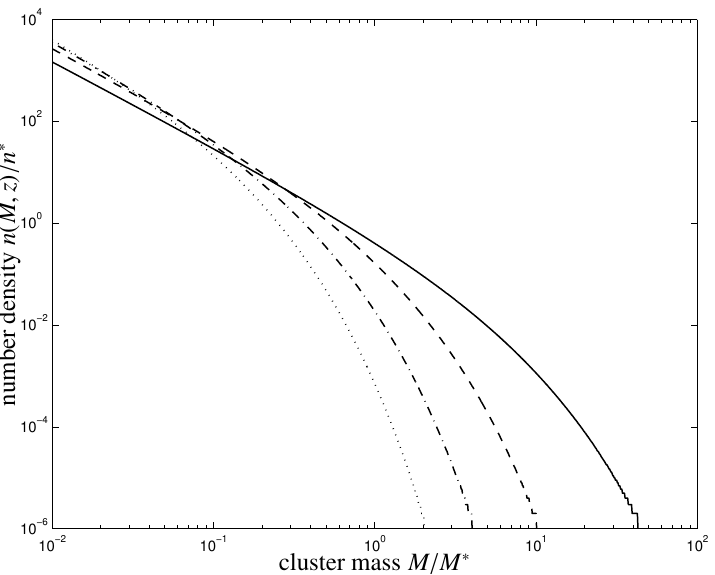}}
\end{center}
\caption{Mass function of haloes as a function of redshift, for $z=0$ (solid line), $z=1$ (dashed line), $z=2$ (dash-dotted line) and $z=4$ (dotted line).}
\label{fig_collapse}
\end{figure}

The number density of haloes of a certain mass can be derived from the fluctuation statistics of a random field \citep{1986ApJ...304...15B}: The cumulative distribution function $P_\delta(M)$ of the amplitudes $\delta$ gives the probability that at a given point is larger than $\delta$, 
\begin{equation}
P_\delta(M) = \int_{\delta_c}^\infty\dd\delta\:p(\delta)
\end{equation}
where the variance of the field is obtained through an integration over the spectrum of the field, smoothed on a spatial scale which reflects the mass-scale of the halo,
\begin{equation}
\sigma^2_R = \int\frac{k^2\dd k}{2\pi^2}\:P(k) W(kR)^2
\end{equation}
with a smoothing function $W(kR)$ that can be chosen to be a top-hat or a Gaussian smoothing kernel. The relationship between $R$ and $M$ is given by $M = 4\pi/3\:\Omega_m\rho_\mathrm{crit}R^3$. The distribution of the halo masses is then given by the derivative $\partial P/\partial M$. For a Gaussian distribution of amplitudes of the density field one obtains
\begin{equation}
n(M) = \frac{\partial P}{\partial M} = \sqrt{\frac{2}{\pi}} \frac{\delta_c}{\sigma_RD_+}\frac{\dd\ln\sigma_R}{\dd M} \exp\left(-\frac{\delta_c^2}{\sigma_R^2D_+^2}\right)
\end{equation}
with the corrected normalisation. The mass function $n(M)$ can be approximated by a power-law with an exponential cutoff in mass,
\begin{equation}
\frac{n(M,z)}{n_*} = \frac{1}{\sqrt{\pi}}\left(1+\frac{n}{3}\right) \frac{\bar{\rho}}{M^2}
\left(\frac{M}{M^*}\right)^{(3+)/6}\exp\left(-\left(\frac{M}{M^*}\right)^{(3+n)/3}\right)
\end{equation}
where the mass scale $M^*$ is a decreasing function with redshift $z$, and $\bar{\rho} = \Omega_M\rho_\mathrm{crit}$ is the background density. Fig.~\ref{fig_collapse} gives an impression of the evolution of the mass function with time. In this specific formula, a scale-free CDM-spectrum of the type $P(k)\propto k^n$ was assumed, which allows a direct computation of $\sigma_R^2$ and its derivatives for specific choices of the filter function.

\subsection{Tidal torquing and the generation of angular momentum}
The angular momentum of a halo follows from the integration of the momentum density over the physical volume of the halo weighted by the distance of each mass element from the centre of gravity
\begin{equation}
\mathbf{L}(\eta) = \int_V\dd^3r\:\left(\vecr-\bar{\vecr}\right)\times\vecv(\vecr,\eta)\rho(\vecr,\eta),
\end{equation}
where $\vecv(\vecr,\eta)$ is velocity of the mass element with density $\rho(\vecr,\eta) = \bra\rho\ket(1+\delta(\vecr,\eta))$ at position $\vecr$ around the centre of gravity $\bar{\vecr}$. 

Simplifying this relationship by assuming a constant density over the volume of the protohalo $\bra\rho\ket = \Omega_m\rho_\mathrm{crit}$ inside the protogalactic region, \citet{1984ApJ...286...38W}, \citet{1996MNRAS.282..436C}, \citet{1997ASPC..117..431T} and \citet{2001ApJ...559..552C} arrive at a concise expression by describing the particle motion in the Zel'dovich approximation:
\begin{equation}
\vecx(\vecq,t) = \vecq - D_+(\eta)\nabla\Psi(\vecq)
\rightarrow
\dot{\vecx} = -\dot{D}_+\nabla\Psi,
\end{equation}
where the dot denotes a derivative with respect to cosmic time $t$. In the centre of mass-frame the expression for the angular momentum becomes
\begin{equation}
\mathbf{L}
= \rho_0 a^5 \int_{V_L}\dd^3q\:\left(\vecx-\bar{\vecx}\right)\times \dot{\vecx}
\simeq \rho_0 a^5 \int_{V_L}\dd^3q \left(\vecq - \bar{\vecq}\right)\times \dot{\vecx},
\end{equation}
when switching from physical to comoving coordinates.

The displacement field $\nabla\Psi(\vecq)$ can be Taylor-expanded in the vicinity of the centre of mass $\bar{\vecq}$
\begin{equation}
\partial_\alpha\Psi(\vecq) \simeq \partial_\alpha\Psi(\bar{\vecq}) + \sum_\beta (\vecq - \bar{\vecq})_\beta\Psi_{\alpha\beta}, 
\end{equation}
such that the tidal shear
\begin{equation}
\Psi_{\sigma\gamma}(\bar{\vecq}) = \partial_\sigma\partial_\gamma\Psi(\bar{\vecq}),
\end{equation}
becomes the one of the central quantities responsible for angular momentum generation. Identifying the tensor of second moments of the mass distribution of the protogalactic object as the moment of inertia $I_{\beta\sigma}$,
\begin{equation}
I_{\beta\sigma} = \rho_0 a^3 \int_{V_L}\dd^3q\: (\vecq - \bar{\vecq})_\beta (\vecq -\bar{\vecq})_\sigma
\end{equation}
one obtains the final expression of the angular momentum $L_\alpha$:
\begin{equation}
L_\alpha = a^2 \dot{D}_+ \epsilon_{\alpha\beta\gamma} \sum_\sigma I_{\beta\sigma}\Psi_{\sigma\gamma}.
\end{equation}

Physically this corresponds to the case when inertia and tidal shear do not share a common eigensystem: The product $X$ of inertia $I_{\beta\sigma}$ and tidal shear $\Psi_{\sigma\gamma}$ can be decomposed into its symmetric part $X^+_{\beta\gamma} = (I_{\beta\sigma}\Psi_{\sigma\gamma} + \Psi_{\beta\sigma}I_{\sigma\gamma})/2 = \left\{I,\Psi\right\}_{\beta\gamma}/2$ and into the corresponding antisymmetric part $X^-_{\beta\gamma} = (I_{\beta\sigma}\Psi_{\sigma\gamma} - \Psi_{\beta\sigma}I_{\sigma\gamma})/2 = \left[I,\Psi\right]_{\beta\gamma}/2$. The contraction with $\epsilon_{\alpha\beta\gamma}$ will be proportional to $X^-$, i.e. the commutator between the shear and inertia tensors. Hence, angular momentum is only generated if those two tensors do not share a common eigensystem. Clearly, no object would start rotating in a spherically symmetric potential, and one needs a misalignment between the two eigensystems for tidal torquing to be effective.

\section{Open questions}\label{bjoern_questions}
The description of cosmic structure formation worked under the assumption of fluid mechanics, which is a questionable concept in the case of dark matter: Dark matter particles have a very small cross section for elastic collisions and therefore, systems of dark matter will not exibit properties like pressure and viscosity whose microscopic origin are particle collisions. A significant progress in this direction has been made by employing statistical field theory \citep{bartelmann_microscopic_2014}.

At the stage when the cosmic density field reaches large amplitudes all perturbative approaches fail: This is the regime of numerical simulations, which have reached an incredible level of sophistication and go in their description much beyond dark matter dynamics, by including a multiphase baryonic component and by tracking its evolution in terms of temperature and chemical composition. How exactly the baryonic component reacts back onto the dark matter component through gravitational forces is still unclear, likewise the exact thermal evolution of the baryonic component is still a matter of research, and different feedback mechanisms are investigated that keep star formation at an acceptably low level. There are, in addition, a number of observations with are being debated, for instance the abundance of subhaloes in dark matter structures and large bulk flows. Whether they are indicative of a failure of the assumption of cold dark matter or of a different gravitational theory, is yet unclear. And it should be emphasised that algorithmic advances in the discretisation have solved problems which were thought to be of physical origin, for instance the very thin discs of spiral galaxies which were previously not reconcilable with numerical simulations \citep{springel_e_2010}.

On larger scales one hopes to investigate deviations of the gravitational law from general relativity by measuring the structure formation rate on large scales, which would provide a test of gravity in a weak-field, slow-motion and large-scale limit. And lastly, it is generally accepted without a detailed proof that the structured metric on small scales transitions to the smooth metric with the FLRW-symmetries on larger scales, and that the effects of the averaging process are small and not compatible with the phenomenology of dark energy.

\section{Summary}\label{bjoern_summary}
The gravitational dynamics of the Universe is governed by general relativity, which is the most general metric theory in gravity in four dimensions with a second-order local and energy-momentum conserving field equation. Assuming spatial homogeneity and isotropy leads to the Friedmann-Lema{\^i}tre-cosmologies, which link the expansion dynamics of the Universe to the density of the gravitating substances and the cosmological constant. The expansion dynamics is well described by a FLRW-model with no spatial curvature, with a matter density $\Omega_m\simeq0.3$ and a cosmological constant whose value corresponds to $\Omega_\Lambda\simeq 1-\Omega_m$.

Cosmic inflation is a physical mechanism for pushing spatial curvature to very small values, and for generating fluctuations in the distribution of (predominantly dark) matter. These structures seeded in the early universe grew by self-gravity. Cosmic structure formation is influenced by the time evolution of the background cosmological model and shows for small amplitudes a straightforward solution in terms of the growth function $D_+$. In this limit, the growth is homogeneous and conserves all statistical properties of the density field. In addition, the velocity field is a gradient field whose divergence is proportional to the density field. 

When the amplitudes in the density field become large, the growth turns nonlinear and generates new, in particular non-Gaussian statistical properties, through the mode-coupling mechanism. 

Isolated peaks in the density field collapse under their own gravity and form haloes, which ultimately host galaxies. Their number density can be computed from the fluctuation statistics of a random field, which in cosmology is in its statistical properties close to Gaussian. Tidal fields can introduce angular momentum into aspherical haloes up to the moment of gravitational collapse.

\bibliographystyle{ws-rv-har}
\bibliography{references}

\begin{thebibliography}{41}
\newcommand{\enquote}[1]{#1}
\providecommand{\natexlab}[1]{#1}
\providecommand{\url}[1]{\texttt{#1}}
\providecommand{\urlprefix}{URL }
\expandafter\ifx\csname urlstyle\endcsname\relax
  \providecommand{\doi}[1]{doi:\discretionary{}{}{}#1}\else
  \providecommand{\doi}{doi:\discretionary{}{}{}\begingroup
  \urlstyle{rm}\Url}\fi

\bibitem[{{Abramowitz} and {Stegun}(1972)}]{1972hmf..book.....A}
{Abramowitz}, M. and {Stegun}, I.~A. (1972). \emph{{Handbook of Mathematical
  Functions}}, Handbook of Mathematical Functions, New York: Dover, 1972.

\bibitem[{{Arfken} and {Weber}(2005)}]{2005mmp..book.....A}
{Arfken}, G.~B. and {Weber}, H.~J. (2005). \emph{{Mathematical methods for
  physicists 6th ed.}}, Materials and Manufacturing Processes.

\bibitem[{{Bardeen} \emph{et~al.}(1986){Bardeen}, {Bond}, {Kaiser} and
  {Szalay}}]{1986ApJ...304...15B}
{Bardeen}, J.~M., {Bond}, J.~R., {Kaiser}, N. and {Szalay}, A.~S. (1986).
  \enquote{{The statistics of peaks of Gaussian random fields},} \emph{\apj}
  \textbf{304}, pp. 15--61, \doi{10.1086/164143}.

\bibitem[{Bartelmann \emph{et~al.}()Bartelmann, Fabis, Berg, Kozlikin, Lilow
  and Viermann}]{bartelmann_microscopic_2014}
Bartelmann, M., Fabis, F., Berg, D., Kozlikin, E., Lilow, R. and Viermann, C.
  (????). \enquote{A microscopic, non-equilibrium, statistical field theory for
  cosmic structure formation,}  \urlprefix\url{http://arxiv.org/abs/1411.0806}.

\bibitem[{{Baumann} and {Peiris}(2008)}]{2008arXiv0810.3022B}
{Baumann}, D. and {Peiris}, H.~V. (2008). \enquote{{Cosmological Inflation:
  Theory and Observations},} \emph{ArXiv e-prints 0810.3022} .

\bibitem[{{Bernardeau} \emph{et~al.}(2002){Bernardeau}, {Colombi},
  {Gazta{\~n}aga} and {Scoccimarro}}]{2002PhR...367....1B}
{Bernardeau}, F., {Colombi}, S., {Gazta{\~n}aga}, E. and {Scoccimarro}, R.
  (2002). \enquote{{Large-scale structure of the Universe and cosmological
  perturbation theory},} \emph{\physrep} \textbf{367}, pp. 1--3.

\bibitem[{{Bernardeau} \emph{et~al.}(2008){Bernardeau}, {Crocce} and
  {Scoccimarro}}]{2008PhRvD..78j3521B}
{Bernardeau}, F., {Crocce}, M. and {Scoccimarro}, R. (2008).
  \enquote{{Multipoint propagators in cosmological gravitational instability},}
  \emph{\prd} \textbf{78}, 10, pp. 103521--+, \doi{10.1103/PhysRevD.78.103521}.

\bibitem[{{Boerner}(2003)}]{2003euff.book.....B}
{Boerner}, G. (2003). \emph{{The early universe : facts and fiction}}, The
  early universe : facts and fiction, 4th ed.~By G.~Boerner.~ Astronomy and
  astrophysics library.~Berlin: Springer, 2003.

\bibitem[{{Bouchet} \emph{et~al.}(1992){Bouchet}, {Juszkiewicz}, {Colombi} and
  {Pellat}}]{1992ApJ...394L...5B}
{Bouchet}, F.~R., {Juszkiewicz}, R., {Colombi}, S. and {Pellat}, R. (1992).
  \enquote{{Weakly nonlinear gravitational instability for arbitrary Omega},}
  \emph{\apjl} \textbf{394}, pp. L5--L8, \doi{10.1086/186459}.

\bibitem[{{Buchert}(1989)}]{1989A&A...223....9B}
{Buchert}, T. (1989). \enquote{{A class of solutions in Newtonian cosmology and
  the pancake theory},} \emph{\aap} \textbf{223}, pp. 9--24.

\bibitem[{{Buchert}(1992)}]{1992MNRAS.254..729B}
{Buchert}, T. (1992). \enquote{{Lagrangian theory of gravitational instability
  of Friedman-Lemaitre cosmologies and the 'Zel'dovich approximation'},}
  \emph{\mnras} \textbf{254}, pp. 729--737.

\bibitem[{{Buchert}(1994)}]{1994MNRAS.267..811B}
{Buchert}, T. (1994). \enquote{{Lagrangian Theory of Gravitational Instability
  of Friedman-Lemaitre Cosmologies - a Generic Third-Order Model for Nonlinear
  Clustering},} \emph{\mnras} \textbf{267}, pp. 811--+.

\bibitem[{{Buchert} and {G{\"o}tz}(1987)}]{1987JMP....28.2714B}
{Buchert}, T. and {G{\"o}tz}, G. (1987). \enquote{{A class of solutions for
  self-gravitating dust in Newtonian gravity},} \emph{Journal of Mathematical
  Physics} \textbf{28}, pp. 2714--2719.

\bibitem[{{Catelan}(1995)}]{1995MNRAS.276..115C}
{Catelan}, P. (1995). \enquote{{Lagrangian dynamics in non-flat universes and
  non-linear gravitational evolution},} \emph{\mnras} \textbf{276}, pp.
  115--124.

\bibitem[{{Catelan} \emph{et~al.}(1995){Catelan}, {Lucchin}, {Matarrese} and
  {Moscardini}}]{1995MNRAS.276...39C}
{Catelan}, P., {Lucchin}, F., {Matarrese}, S. and {Moscardini}, L. (1995).
  \enquote{{Eulerian perturbation theory in non-flat universes: second-order
  approximation},} \emph{\mnras} \textbf{276}, pp. 39--56.

\bibitem[{{Catelan} and {Theuns}(1996)}]{1996MNRAS.282..436C}
{Catelan}, P. and {Theuns}, T. (1996). \enquote{{Evolution of the angular
  momentum of protogalaxies from tidal torques: Zel'dovich approximation},}
  \emph{\mnras} \textbf{282}, pp. 436--454.

\bibitem[{{Cheng}(2005)}]{2005rgc..book.....C}
{Cheng}, T.-P. (2005). \emph{{Relativity, gravitation and cosmology. A basic
  introduction}}.

\bibitem[{{Chevallier} and {Polarski}(2001)}]{2001IJMPD..10..213C}
{Chevallier}, M. and {Polarski}, D. (2001). \enquote{{Accelerating Universes
  with Scaling Dark Matter},} \emph{International Journal of Modern Physics D}
  \textbf{10}, pp. 213--223, \doi{10.1142/S0218271801000822}.

\bibitem[{{Crittenden} \emph{et~al.}(2001){Crittenden}, {Natarajan}, {Pen} and
  {Theuns}}]{2001ApJ...559..552C}
{Crittenden}, R.~G., {Natarajan}, P., {Pen}, U.-L. and {Theuns}, T. (2001).
  \enquote{{Spin-induced Galaxy Alignments and Their Implications for
  Weak-Lensing Measurements},} \emph{\apj} \textbf{559}, pp. 552--571,
  \doi{10.1086/322370}.

\bibitem[{{Crocce} and {Scoccimarro}(2006{\natexlab{a}})}]{2006PhRvD..73f3520C}
{Crocce}, M. and {Scoccimarro}, R. (2006{\natexlab{a}}). \enquote{{Memory of
  initial conditions in gravitational clustering},} \emph{\prd} \textbf{73}, 6,
  pp. 063520--+, \doi{10.1103/PhysRevD.73.063520}.

\bibitem[{{Crocce} and {Scoccimarro}(2006{\natexlab{b}})}]{2006PhRvD..73f3519C}
{Crocce}, M. and {Scoccimarro}, R. (2006{\natexlab{b}}). \enquote{{Renormalized
  cosmological perturbation theory},} \emph{\prd} \textbf{73}, 6, pp.
  063519--+, \doi{10.1103/PhysRevD.73.063519}.

\bibitem[{{Durrer}(2008)}]{2008cmbg.book.....D}
{Durrer}, R. (2008). \emph{{The Cosmic Microwave Background}}.

\bibitem[{{Ehlers} and {Buchert}(1997)}]{1997GReGr..29..733E}
{Ehlers}, J. and {Buchert}, T. (1997). \enquote{{Newtonian Cosmology in
  Lagrangian Formulation: Foundations and Perturbation Theory},} \emph{General
  Relativity and Gravitation} \textbf{29}, pp. 733--764.

\bibitem[{{Eisenstein} and {Hu}(1998)}]{1998ApJ...496..605E}
{Eisenstein}, D.~J. and {Hu}, W. (1998). \enquote{{Baryonic Features in the
  Matter Transfer Function},} \emph{\apj} \textbf{496}, pp. 605--+,
  \doi{10.1086/305424}.

\bibitem[{{Eisenstein} and {Hu}(1999)}]{1999ApJ...511....5E}
{Eisenstein}, D.~J. and {Hu}, W. (1999). \enquote{{Power Spectra for Cold Dark
  Matter and Its Variants},} \emph{\apj} \textbf{511}, pp. 5--15,
  \doi{10.1086/306640}.

\bibitem[{{Hobson} \emph{et~al.}(2006){Hobson}, {Efstathiou} and
  {Lasenby}}]{2006gere.book.....H}
{Hobson}, M.~P., {Efstathiou}, G.~P. and {Lasenby}, A.~N. (2006).
  \emph{{General Relativity}}, \doi{10.2277/0521829518}.

\bibitem[{{Langlois}(2010)}]{2010arXiv1001.5259L}
{Langlois}, D. (2010). \enquote{{Lectures on inflation and cosmological
  perturbations},} \emph{ArXiv e-prints 1001.5259} .

\bibitem[{{Liddle} and {Lyth}(2000)}]{2000cils.book.....L}
{Liddle}, A.~R. and {Lyth}, D.~H. (2000). \emph{{Cosmological Inflation and
  Large-Scale Structure}}, Cosmological Inflation and Large-Scale Structure, by
  Andrew R.~Liddle and David H.~Lyth, pp.~414.~ISBN 052166022X.~Cambridge, UK:
  Cambridge University Press, April 2000.

\bibitem[{{Linder}(2005)}]{2005PhRvD..72d3529L}
{Linder}, E.~V. (2005). \enquote{{Cosmic growth history and expansion
  history},} \emph{\prd} \textbf{72}, 4, pp. 043529--+,
  \doi{10.1103/PhysRevD.72.043529}.

\bibitem[{{Linder} and {Jenkins}(2003)}]{2003MNRAS.346..573L}
{Linder}, E.~V. and {Jenkins}, A. (2003). \enquote{{Cosmic structure growth and
  dark energy},} \emph{\mnras} \textbf{346}, pp. 573--583,
  \doi{10.1046/j.1365-2966.2003.07112.x}.

\bibitem[{{Longair}(2008)}]{2008gafo.book.....L}
{Longair}, M.~S. (2008). \emph{{Galaxy Formation}}, Galaxy Formation, by
  Malcolm S.~Longair Berlin: Springer, 2008.~ ISBN 978-3-540-73477-2.

\bibitem[{{Peacock}(1999)}]{1999coph.book.....P}
{Peacock}, J.~A. (1999). \emph{{Cosmological Physics}}, Cosmological Physics,
  by John A.~Peacock, pp.~704.~ISBN 052141072X.~Cambridge, UK: Cambridge
  University Press, January 1999.

\bibitem[{{Peebles}(1980{\natexlab{a}})}]{1980PhyS...21..720P}
{Peebles}, P.~J.~E. (1980{\natexlab{a}}). \enquote{{Nature of the matter
  distribution now and at Z = 1000},} \emph{\physscr} \textbf{21}, pp.
  720--724.

\bibitem[{{Peebles}(1980{\natexlab{b}})}]{1980lssu.book.....P}
{Peebles}, P.~J.~E. (1980{\natexlab{b}}). \emph{{The large-scale structure of
  the universe}}.

\bibitem[{{Sahni} and {Coles}(1995)}]{1995PhR...262....1S}
{Sahni}, V. and {Coles}, P. (1995). \enquote{{Approximation methods for
  non-linear gravitational clustering},} \emph{\physrep} \textbf{262}, pp.
  1--135.

\bibitem[{Springel()}]{springel_e_2010}
Springel, V. (????). \enquote{E pur si muove: {{Galiliean}}-invariant
  cosmological hydrodynamical simulations on a moving mesh,}  \textbf{401}, 2,
  pp. 791--851, \doi{10.1111/j.1365-2966.2009.15715.x},
  \urlprefix\url{http://arxiv.org/abs/0901.4107}.

\bibitem[{{Sugiyama}(1995)}]{1995ApJS..100..281S}
{Sugiyama}, N. (1995). \enquote{{Cosmic Background Anisotropies in Cold Dark
  Matter Cosmology},} \emph{\apjs} \textbf{100}, pp. 281--+,
  \doi{10.1086/192220}.

\bibitem[{{Theuns} and {Catelan}(1997)}]{1997ASPC..117..431T}
{Theuns}, T. and {Catelan}, P. (1997). \enquote{{Angular Momentum Induced by
  Tidal Torques},} in M.~{Persic} and P.~{Salucci}. eds., \emph{Dark and
  Visible Matter in Galaxies and Cosmological Implications}, \emph{Astronomical
  Society of the Pacific Conference Series}, Vol. 117, pp. 431--+.

\bibitem[{{Turner} and {White}(1997)}]{1997PhRvD..56.4439T}
{Turner}, M.~S. and {White}, M. (1997). \enquote{{CDM models with a smooth
  component},} \emph{\prd} \textbf{56}, pp. 4439--+.

\bibitem[{{White}(1984)}]{1984ApJ...286...38W}
{White}, S.~D.~M. (1984). \enquote{{Angular momentum growth in protogalaxies},}
  \emph{\apj} \textbf{286}, pp. 38--41, \doi{10.1086/162573}.

\bibitem[{{Zel'dovich}(1970)}]{1970A&A.....5...84Z}
{Zel'dovich}, Y.~B. (1970). \enquote{{Gravitational instability: An approximate
  theory for large density perturbations.}} \emph{\aap} \textbf{5}, pp. 84--89.

\end{thebibliography}

\end{document}